\documentclass{ISMA_USD}

\usepackage{soul}
\usepackage{xcolor}
\usepackage{arydshln}
\definecolor{mainblue}{RGB}{0,70,140}

\hypersetup{pdftitle = {Acceleration data-based robust on-board monitoring of hunting proximity in railway vehicles under uncertainty: A feasibility study},
    pdfauthor = {K. Kritikakos, S. D. Fassois, J. S. Sakellariou},
    pdfkeywords = {railway hunting, hunting proximity, maximum Lyapunov exponent, grey model}}

\title{Acceleration data-based robust on-board monitoring of \\hunting proximity in railway vehicles under uncertainty:\\ A feasibility study}

\author[]{K. Kritikakos}
\author[]{S.D. Fassois}
\author[]{J.S. Sakellariou}
		
\affil{University of Patras, Department of Mechanical Engineering \& Aeronautics, Stochastic Mechanical Systems \& Automation (SMSA) Laboratory, \\ 26504 Patras, Greece 
	 \NewLineAffil e-mail: \textbf{fassois@upatras.gr}}

\date{}

\usepackage[numbers,sort&compress]{natbib}

\hypersetup{
    colorlinks=true,
    linkcolor=mainblue,
    citecolor=mainblue,
    urlcolor=mainblue
}

\begin{document}
%\bstctlcite{IEEEtran:BSTadapt}	

%%%%%%%%%%%%
% Abstract %
%%%%%%%%%%%%

% Enter your abstract in the command below.

\abstract{
The feasibility of on-board, real-time proximity assessment to railway hunting initiation under uncertainty arising from vehicle/track and other operational conditions is assessed. In this context, a novel method that classifies hunting proximity into one of several prespecified speed proximity intervals is introduced, based on a grey nonlinear vehicle model with parameters estimated from on-board acceleration measurements and model-based estimation of the Maximum Lyapunov Exponent (MLE) as a hunting proximity measure.
The method's performance and robustness are assessed via Monte Carlo simulations using a MATLAB-based multi-body vehicle model and four operating scenarios. The method is shown to achieve excellent performance achieving 97.08\% classification accuracy, which marks a significant improvement over a potential alternative. Its robustness to faulty suspensions and hollow-worn wheels is also demonstrated, although certain performance compromise is observed under worn-track conditions.}

\maketitle

%%%%%%%%%%%%%%%%%%%%%%%%%%
% Beginning of the paper %
%%%%%%%%%%%%%%%%%%%%%%%%%%

\section{Introduction}

Railway hunting is a phenomenon of critical importance in the railway industry for reasons associated with safety, ride comfort, and rolling stock life cycle \cite{WICKENS1998}. Hunting is the coupled lateral and yaw limit cycle (periodic) oscillation of the vehicle, physically limited by the wheel flanges \cite{WICKENS1998}.
It occurs at a traveling speed $v_h$, referred to as the \emph{hunting speed}, which depends on the specific track/vehicle conditions \cite{KRITIKAKOS2024}. This phenomenon fundamentally arises from the nonlinear nature of railway vehicle dynamics, mainly associated with wheel–rail contact kinematics, contact forces, and suspension system components \cite[pp. 206–207]{GARG1984}.

The related problem of on-board hunting onset \emph{detection} (following hunting initiation) has attracted considerable attention \cite{SUN2021, GUO2024,Kulkarni2025}. Yet, a key limitation is that robustness to uncertainty arising from vehicle/track condition and other operational factors - which is instrumental for avoiding false alarms - has been scarcely addressed. To fill this gap, the present authors have introduced robust hunting detection methods based on adaptive stochastic parametric modeling and innovative hunting indicators \cite{KRITIKAKOS2025,KRITIKAKOS2024,KRITIKAKOS2022,KRITIKAKOS2021}. These have been shown to achieve excellent performance (100\% correct detection for 0\% false alarm), with a mean detection time delay of 0.2 s and robustness to various track/vehicle conditions. 

The central \ul{question} raised in this study is whether the instantaneous proximity to hunting onset can be robustly and reliably evaluated during vehicle operation under any operating conditions and prior to hunting occurrence. To this end, hunting proximity assessment is treated as a classification problem using several prespecified traveling speed intervals. An innovative method aiming at effectiveness and robustness is introduced based on: (i) A grey nonlinear model of the vehicle’s lateral (and thus hunting-related) dynamics, incorporating a physics-informed structure with unknown parameters estimated from on-board acceleration measurements; and (ii) grey-model-based estimation of the Maximum Lyapunov Exponent (MLE) \cite{ABARBANEL1993,PARLITZ2016} as a measure of hunting proximity.

The MLE is widely employed in the analysis of stability and chaotic behavior in nonlinear systems \cite{ABARBANEL1993,PARLITZ2016}. It characterizes whether two response trajectories of a system, starting at arbitrarily close initial conditions, converge (negative MLE), maintain a constant separation (as in a limit cycle; zero MLE), or diverge (positive MLE) \cite{ABARBANEL1993}. Since railway hunting is a limit cycle oscillation, the MLE is expected to be zero or become slightly positive when weak chaotic components are induced by wheel–rail flange contact \cite{TRUE1992}. Thus, as the vehicle speed $v$ approaches the condition-dependent and generally unknown hunting speed $v_h$, the MLE is expected to evolve from negative values toward zero or slightly positive values, independently of vehicle and track conditions.

MLE computation is conceptually straightforward when the underlying dynamical equations are known; however, these are rarely available in practical applications \cite{ABARBANEL1993,PARLITZ2016}. This has led to data-driven estimation methods based solely on measured system responses, broadly classified as \emph{direct} \cite{BRARI2022,ROSENSTEIN1993,WOLF1985} and \emph{indirect} \cite{YAKOVLEVA2025,ESCOT2023,PATHAK2017}. \emph{Direct methods} estimate the MLE by identifying nearby state vectors within the measured responses and evaluating how their separation evolves over time. 
Although conceptually simple, these methods are sensitive to measurement noise and external disturbances, which may induce artificial state-vector divergence and false indications of chaotic behavior, while in high-order systems ($n\geq 10$) the number of samples required to identify sufficiently close neighboring state vectors rapidly increases with system dimension \cite{YAKOVLEVA2025,ABARBANEL1993,PARLITZ2016,WOLF1985}.
On the contrary, \emph{indirect methods} are based on estimating a nonlinear regression model from the measured responses, differentiating the model to obtain its Jacobian, and then applying standard MLE computation techniques, as in the known-dynamics case.
Typical model structures involve polynomial-type \citep{ESCOT2023} and Neural Network (NN) \citep{YAKOVLEVA2025,ESCOT2023,PATHAK2017} type approximations. In contrast to direct methods, this category of methods generally copes well with signal randomness \citep{ESCOT2023}; however, for high-order systems the accuracy of the estimated nonlinear model deteriorates. As the model order $n$ increases, the convergence of the model derivatives (and thus the resulting Jacobian matrix) deteriorates, and differentiation amplifies modeling errors that accumulate during MLE computation \citep{PARLITZ2016}. %In addition, both categories generally require careful tuning of several critical design parameters.

Furthermore, railway vehicles are characterized by complex and high-order dynamics ($n\geq34$), implying that accurate MLE estimation using existing direct or indirect methods becomes highly challenging. To address this, a dedicated grey nonlinear railway vehicle model with a physics-informed structure is presently developed to represent the vehicle’s lateral, and consequently hunting-related, dynamics. The model parameters, assumed unknown, are estimated via Linear Least Squares (LLS) \citep[p. 206]{LJUNG1999} type criteria using acceleration measurements from multiple uni-axial onboard sensors.  
Using the Jacobian matrix of the estimated grey model, the MLE can be computed using available techniques \citep[pp. 74–80]{PARKER1989} and is subsequently employed to assess hunting proximity by classifying the proximity measure $\Delta v$ - representing the difference between the current vehicle speed $v$ and the unknown, condition-dependent hunting speed $v_h$ - into one of several user-defined proximity intervals $\Delta v_i$ ($i=1,\ldots,d$).

The method's performance is systematically assessed via numerous Monte Carlo simulations using a multi-body vehicle model developed in MATLAB and four realistic operating scenarios corresponding to different suspension, wheel and track conditions. For each operating scenario, several Monte Carlo simulations are performed at constant speeds $v$, randomly selected within each considered hunting proximity interval $\Delta v_{i}$ ($i=1,\ldots,d$). The objective is to correctly assess the proximity to hunting onset under uncertainty arising from track and vehicle conditions that lead to different (and typically unknown) hunting speeds $v_h$.
Performance assessment evaluation is based on confusion matrices, as well as, on the Accuracy (ACC) metric which represents the correct classification rate. For a preliminary version of the postulated method the reader is referred to our recent study \cite{KRITIKAKOS2024_CONF}.

The main \emph{original contributions} of the study may be summarized as follows:
\begin{enumerate}\vspace*{-2mm}
    \item[1.] The feasibility of assessing the proximity to the hunting onset during normal vehicle operation is examined for the first time, and an innovative MLE-based method is proposed for this purpose. Such capability is of major potential practical importance, as it can support higher travel speeds while maintaining safe margins from hunting. \vspace*{-2mm}
    \item[2.] A grey-model-based MLE estimation approach is proposed to overcome limitations of state-of-the-art data-driven MLE methods when applied to high-order, and complex systems such as railway vehicles. \vspace*{-6mm} %The grey model adopts a physics-informed structure with parameters identified from on-board measurements during normal operation, providing an accurate representation of nonlinear lateral (hunting-related) dynamics and enabling reliable MLE estimation. \vspace*{-3mm} %The explicit representation of complex wheel–rail contact nonlinearities, together with input (track irregularity) estimation, constitutes key features of the proposed grey model.\vspace*{-3mm}
    \item[3.] Critical comparisons with an additionally postulated method that is based on state-of-the-art indirect (data-driven) MLE estimation \cite{ESCOT2023} properly adapted to non-autonomous systems. \vspace*{-2mm}
\end{enumerate} 

The rest of this paper is structured as follows: The precise problem statement is presented in \autoref{S.2}, while details on the Monte Carlo simulations are provided in \autoref{S.3}. The hunting proximity assessment method is introduced in \autoref{S.4}, its performance is assessed in \autoref{S.5} and conclusions are drawn in \autoref{S.6}.

%-------------------------------------------------------------------------
\section{Precise problem statement}\label{S.2}
The problem of hunting proximity assessment (classification) may be stated as follows:
Given a $N$-sample-long vector acceleration signal $\bm{y}[k]$ ($k=1,\ldots,N$), available at time instant $N$ through $s$ uni-axial on-board sensors while the vehicle travels at a constant (over the observation interval) speed $v$, \emph{properly} classify the proximity of the current speed $v$ to the (unknown and condition-dependent) hunting speed $v_h$ into one of $d$ user-defined intervals $\Delta v_i$ ($i=1,\ldots,d$). The procedure may be repeated at any  subsequent time instant.

\noindent  \underline{Remarks:} 
    \begin{enumerate}\vspace*{-2mm}
    \item[1.] In the above $k$ within brackets designating the discrete time normalized by the sampling period $T_s$; with continuous time being later on designated as $t$ within parentheses. \vspace*{-3mm}
    \item[2.] The term `properly' refers to the hunting proximity being assessed correctly and independently of vehicle and track conditions (\emph{robustness property}).  \vspace*{-3mm}
    \item[3.] In the present study, eight hunting proximity intervals ($d=8$) are considered:  
    \leavevmode
        \begin{equation}\label{eq:Prox_Intervals}
\Delta v_i= (\Delta v_{i,\text{min}},\Delta v_{i,\text{max}}]\;\text{km}/\text{h},\quad i=1,2,\ldots,8 
\end{equation}
with $\Delta v_{i,\text{min}}=10i-5$ and $\Delta v_{i,\text{max}}=10i+5$ $\text{km/h}$. 
    \end{enumerate}

%---------------------------------------------------------
\section{The railway vehicle model and the Monte Carlo simulations}\label{S.3}

\subsection{The railway vehicle simulation model}\label{S.3.1}

\begin{figure}[!b]
\begin{center}
\includegraphics[width=0.9\textwidth,trim={1.33cm 4.15cm 0.2cm 1.7cm},clip]{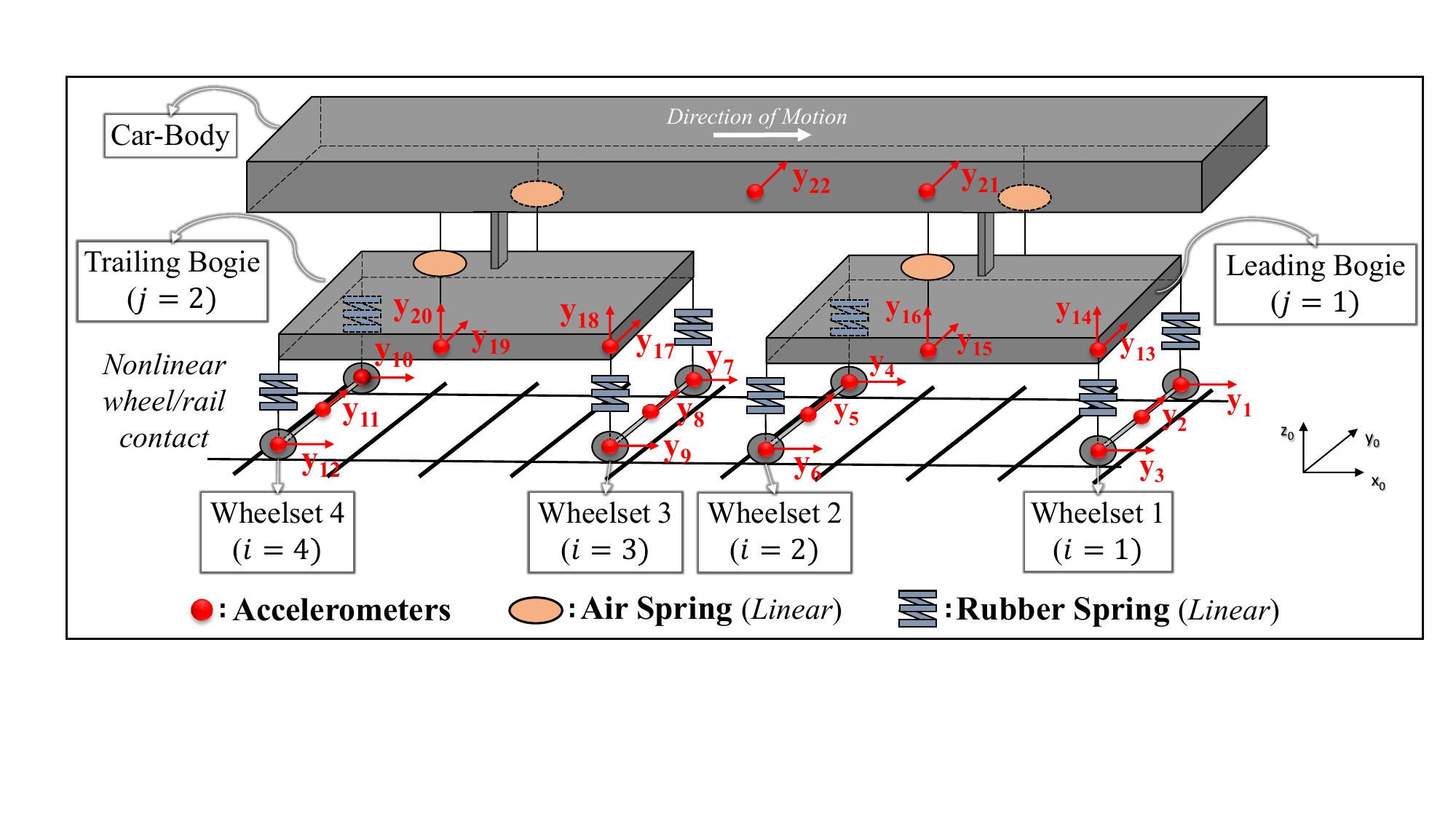} 
\caption{Schematic representation of the vehicle simulation model with the accelerometer locations and measuring directions indicated in red.} 
\label{fig:Simulation_Model_A}
\end{center}
\end{figure}

A 21-degree-of-freedom (DoF) railway vehicle simulation model of the lateral, and thus hunting-related, dynamics is developed in MATLAB and schematically shown in \autoref{fig:Simulation_Model_A}. The model consists of $7$ rigid bodies, including a car-body, two bogie frames, and four wheelsets. Each wheelset is assigned with 3 DoFs corresponding to the lateral deviation $y_{wi}$, longitudinal deviation $x_{wi}$, and yaw angle $\psi_{wi}$, while the roll angle $\phi_{wi}$ is geometrically constrained by wheel–rail contact \cite[pp. 146--147]{GARG1984} ($i=1,\ldots,4$). The bogie frames/car-body are also represented by 3 DoFs, corresponding to lateral deviation $y_{bj}$/$y_{c}$, yaw angle $\psi_{bj}$/$\psi_{c}$ and roll angle $\phi_{bj}$/$\phi_{c}$.  Due to the weak coupling between the vertical and lateral motions of railway vehicles \cite[p. 59]{GARG1984}, and considering that the model's purpose associated with railway hunting, the vertical and pitch motions of all vehicle components, as well the vertical track irregularities, are neglected.

The bogie frames are connected to the wheelsets via conical rubber springs (primary suspension), and the car-body to the bogie frames via air springs (secondary suspension) as shown in \autoref{fig:Simulation_Model_A}. Both are modeled as linear spring-damper elements with stiffness and damping coefficients in the three directions ($x_0,\,y_0,\,z_0$). The following assumptions are adopted in the derivation of the simulation model \cite{GARG1984,IWNICKI2015}: (i) Vehicle motion on a tangent track at constant speed $v$, (ii) each wheel is always in contact with the rail at a single point, (iii) rigid bodies and track, except for elastic deformation in the contact region, (iv) full geometric symmetry about the longitudinal plane, (v) negligible effects of spin creep moments on lateral dynamics, and (vi) linear behavior of the suspension components.

The equations of motion for the $i$-th wheelset, the $j$-th bogie frame and the car-body with respect to its equilibrium coordinate system are given as follows ($i=1,\ldots,4$; $j=1,2$; see also \autoref{fig:Simulation_Model_A}) \cite[p. 76]{Zhai2020}:
\begin{equation}\label{eq:1}
\begin{aligned}
    m_w\Ddot{x}_{wi} ={}& F_{xi}^l + F_{xi}^r + N_{xi}^l + N_{xi}^r - F_{pxi}\\
    m_w\Ddot{y}_{wi} ={}& F_{yi}^l + F_{yi}^r + N_{yi}^l + N_{yi}^r - F_{pyi} - F_{ti} \\
    I_{wz}\Ddot{\psi}_{wi} ={}& -I_{wy}\mathit{\Omega}\Dot{\phi}_{wi} + a_0(F_{xi}^r-F_{xi}^l)+a_0\psi_{wi}(F_{yi}^r+N_{yi}^r-F_{yi}^l-N_{yi}^l)- M_{pzi}\\
    m_t\Ddot{y}_{bj} ={}& F_{py(2j-1)} + F_{py(2j)} - F_{syj} \\
    I_{tz}\Ddot{\psi}_{bj} ={}& M_{pz(2j-1)} + M_{pz(2j)}+l_b(F_{py(2j-1)} - F_{py(2j)}) - d_sF_{sxj} \\
    I_{tx}\Ddot{\phi}_{bj} ={}& h_{w}(F_{py(2j-1)} + F_{py(2j)})+d_w(F_{pz(2j-1)} + F_{pz(2j)})+h_{b}F_{syj} - d_sF_{szj}\\
    m_c\Ddot{y}_{c} ={}& F_{sy1} + F_{sy2} \\
    I_{cz}\Ddot{\psi}_{c} ={}& d_s(F_{sx1} + F_{sx2})+l_c(F_{sy1} + F_{sy2})\\
    I_{cx}\Ddot{\phi}_{c} ={}& h_c(F_{sy1} + F_{sy2})+d_s(F_{sz1} + F_{sz2})\\
\end{aligned} 
\end{equation}
with superscripts $l$ and $r$ designating the left and right wheel respectively, $F_{xi}^{(l,r)}$, $F_{yi}^{(l,r)}$, the longitudinal and lateral creep forces, respectively, $N_{yi}^{(l,r)}$ the lateral component of the normal forces, $F_{pxi}$, $F_{pyi}$, $F_{pzi}$ the longitudinal, lateral and vertical primary suspension forces, respectively, $M_{pzi}$ the  yaw primary suspension moment, $F_{sxj}$, $F_{syj}$, $F_{szj}$ the longitudinal, lateral and vertical secondary suspension forces, $F_{ti}$ the flange contact lateral force, $\mathit{\Omega}$ the wheelset nominal angular velocity. 
The definition of the remaining parameters and their values (based on an Athens Metro vehicle) are provided in \autoref{Table_Model_Pars}. The analytical expressions of the above forces and moments are provided below. 

\vspace{-1 mm}
\noindent \textbf{Wheel-rail creep forces:} By assuming small yaw $\psi_{wi}$, roll $\phi_{wi}$, and contact angles $\delta_i^{(l,r)}$, the creep force components of the wheelset $i$ in its equilibrium coordinate system are calculated based on Kalker's linear theory as follows \citep[pp. 151--154]{GARG1984}:  
\begin{align}\label{eq:2}
    F_{xi}^{(l,r)} ={}& -\varepsilon_i f_{1i}^{(l,r)}\xi_{xi}^{(l,r)} + \varepsilon_i f_{2i}^{(l,r)}\xi_{yi}^{(l,r)}\psi_{wi} +\varepsilon_i f_{3i}^{(l,r)}\xi_{spi}^{(l,r)}\psi_{wi} \\
    F_{yi}^{(l,r)} ={}&  -\varepsilon_i f_{1i}^{(l,r)}\xi_{xi}^{(l,r)} \psi_{wi}  -\varepsilon_i f_{2i}^{(l,r)}\xi_{yi}^{(l,r)} -\varepsilon_i f_{3i}^{(l,r)}\xi_{spi}^{(l,r)}
\end{align}
with $f_{1i}^{(l,r)}$, $f_{2i}^{(l,r)}$, $f_{3i}^{(l,r)}$ designating the creep coefficients, which depend on the wheel-rail contact patch geometry and material properties and obtained as in \cite[p. 133]{Zhai2020}, $\varepsilon_i$ $(:=(\varepsilon_i^l+\varepsilon_i^r)/2)$ the saturation coefficient determined according to the Kalker's empirical law \cite[pp. 259--261]{IWNICKI2015}, and $\xi_{x}^{(l,r)}$, $\xi_{y}^{(l,r)}$, and $\xi_{sp}^{(l,r)}$  the longitudinal, lateral, and spin creepages coefficients calculated as in \citep[p. 154]{GARG1984}.
\vspace{-1 mm}

\noindent \textbf{Suspension forces/moments, normal forces and flange forces:} Both primary and secondary suspension forces/moments are obtained as in  \citep[pp. 72-75]{Zhai2020}. 
The lateral normal-force components $N_{yi}^{(l,r)}$ are calculated as in \citep[p. 157]{GARG1984}, neglecting wheel--rail penetration at the contact patch for simplicity. The flange contact force $F_{ti}$ is modeled as a lateral spring with stiffness $k_{ty}$ and a dead-band $\eta$, corresponding to the flange clearance \cite{BUSTOS2023}.

\vspace{-1 mm}
\noindent\textbf{Wheel-rail contact parameters:} 
Due to the continuously varying curvature of the wheel and rail profiles, the contact parameters including the rolling radius  $r_i^{(l,r)}$, the contact angles $\delta_i^{(l,r)}$, the roll angles $\phi_{wi}$, and the semi-axes of the contact patches $a_{i}^{(l,r)}$, $b_{i}^{(l,r)}$ are nonlinear and non-smooth functions of the wheel excursion $y_{ei}^{(l,r)}=y_{wi} - u_a \pm u_g/2$, where $u_a$ and $u_g$ denote the alignment and gauge-variation track irregularities\footnote{The upper and lower sign in $\pm$ correspond to the left and right wheel, respectively.}, respectively, both being explicitly dependent on time $t$ \citep[pp. 214-215]{GARG1984}.
Under assumptions (ii) and (iv), the wheelset roll angle $\phi_{wi}$ is calculated as: $\phi_{wi}=(r_i^l-r_i^r)/(2a_0)$ \citep[p. 158]{GARG1984}.
During simulations, the wheel-rail contact parameters are obtained from a precomputed look-up table based on the \emph{equivalent-elastic} method \citep{VOLLEBREGT2011}, in which the contact point and associated geometrical parameters are determined through a minimum-distance problem between the undeformed profiles, the contact patch is described through a quasi-Hertzian contact approach, and the creep forces are computed using FASTSIM algorithm.

\vspace{-1 mm}
\noindent \textbf{Track irregularities:} For simplification purposes, in the context of the present study, the gauge variation is assumed to be zero ($u_g=0 \Leftrightarrow y_{ei}^l=y_{ei}^r=y_{ei}$) \cite{WESTON2007}.  Therefore, the simulation model is excited only by the alignment track irregularity $u_a$, which enters the railway vehicle model via the wheel--rail contact parameters \citep[p. 214]{GARG1984} and is modeled as a wide-sense stationary random signal characterized by a spatial Power Spectral Density (PSD) following the ERRI B176 standard \citep{HAIGERMOSER2015}.

\subsection{The operating scenarios and the Monte Carlo simulations}\label{S.3.2}

\begin{table}[!b]
\centering
	\caption{Details on the assessment operating scenarios.}	
	{\begin{tabular}{{>{\centering\arraybackslash}p{3.75cm}
                >{\centering\arraybackslash}p{5.5cm}
                >{\centering\arraybackslash}p{5cm}}}
			\hline
			Operating Scenario & Track/Vehicle Condition & Hunting Speed ($v_h$; in km/h)   \\
			\hline
			\multicolumn{1}{c}{A} &  Nominal & 437.0 \\
			\multicolumn{1}{c}{B} &  Faulty primary suspension & 417.7 \\
			\multicolumn{1}{c}{C} &  Hollow worn wheels &  398.8 \\
			\multicolumn{1}{c}{D} & Worn track  & 428.9 \\
			\hline
            \multicolumn{3}{l}{\small{\emph{Details on scenarios:} Scenario B: 20\% reduction in the stiffness/damping coefficients of the leading bogie}} \\
            \multicolumn{3}{l}{\small{(front left corner; all directions); scenario C: Tread depth $1.48\;$mm (all wheels); scenario D: 50\% increase}} \\
            \multicolumn{3}{l}{\small{in the track irregularity standard deviation above the Alert Limit (AL) value \cite{EN}.}} \\
            \hline
	\end{tabular}}
	\label{Operating_Scenarios}
\end{table}

%---------------------
\begin{table}[!b]
\centering
	\caption{Details on the Monte Carlo simulations and acceleration signals.}	
	{\begin{tabular}{cc}
\hline
 Vehicle Speed ($v$; in km/h)& No. of Simulations per Scenario \\ 
 &  \& Proximity Interval $\Delta v_{i}$  \\ 
\hline
  $ [v_h-\Delta v_{i,\text{max}}+2.5\,\frac{\text{km}}{\text{h}},v_h-\Delta v_{i,\text{min}}-2.5\,\frac{\text{km}}{\text{h}}] \subset \Delta v_{i}$\textsuperscript{(a)}&   15    \\
    ($i=1,2,\ldots,8$) &     \\
\hline
\multicolumn{2}{l}{\small{\emph{Simulation Details}: Integration method: Dormand–Prince \cite{DORMAND1980}; Absolute error tolerance: 1$\times$10\textsuperscript{$-3$}; Relative}} \\
\multicolumn{2}{l}{\small{error tolerance: 1$\times$10\textsuperscript{$-6$}; Maximum step size: 5$\times$10\textsuperscript{$-4$}; MATLAB function: \textcolor{gray}{ode45.m}}}\\ \hdashline
\multicolumn{2}{l}{\small{\emph{Signal Details}: Sampling frequency $f_s=20\,000$ Hz; Signal length $N=1\,600\,000$ samples (80 s).}} \\
\hline
\multicolumn{2}{l}{\small{\textsuperscript{(a)}The proximity intervals $\Delta v_i$ and their bounds, $\Delta v_{i,\text{min}}$ and $\Delta v_{i,\text{max}}$, are defined in \autoref{S.2}.}} \\
\end{tabular}}
	\label{Table:Monte_Carlo_Sims}
\end{table}

The performance of the postulated method is assessed through extensive Monte Carlo simulations conducted under four operating scenarios A–D (see \autoref{Operating_Scenarios}), using the MATLAB-based vehicle model described in \autoref{S.3.1}. For each operating scenario and considered hunting proximity interval $\Delta v_{i}$ ($i=1,\ldots,8$; see \autoref{eq:Prox_Intervals}), Monte Carlo simulations are performed at constant vehicle speeds $v$, randomly selected within the ranges specified in \autoref{Table:Monte_Carlo_Sims}. The objective is to assess the method’s capability to correctly classify the instantaneous proximity to hunting $\Delta v$ ($:= v_h - v$) into predefined intervals $\Delta v_{i}$, under uncertainty induced by varying suspension, track, and wheel conditions that lead to different (and typically unknown) hunting speeds $v_h$.
The hunting speed $v_h$  depends on the track/vehicle design and condition and calculated as in \cite{KRITIKAKOS2024}.
All Monte Carlo simulations are based on random alignment track irregularity profiles (details in \autoref{S.3.1}), resulting in distinct track profile geometries per simulation. For each simulation, 22 random acceleration signals are obtained from accelerometers mounted at locations on the railway vehicle as shown in \autoref{fig:Simulation_Model_A}. %Representative bogie lateral acceleration signals from simulations under nominal operating scenarios and indicative hunting proximity speeds $\Delta v$ are presented in \hyperref[fig:Indicative_Signals]{Figure~\ref{fig:Indicative_Signals}(a)}. At the $\Delta v=0$, or equivalently $v=v_h$, the vehicle clearly enters sustained limit-cycle oscillations characterized by a frequency peak around $3$ Hz, indicating bogie hunting onset (see \hyperref[fig:Indicative_Signals]{Figure~\ref{fig:Indicative_Signals}(b)}). 
Further details on Monte Carlo simulations and resulting acceleration signals are summarized in \autoref{Table:Monte_Carlo_Sims}.

%\begin{figure}[!t]
%\begin{center}
%\includegraphics[width=\textwidth,trim={1.85cm 6.9cm 1.95cm 00cm},clip]%{Images/Indicative_Signals.eps} 
%\caption{Representative simulations under operating scenario A: (a) Bogie lateral acceleration %signals from sensor $15$ under indicative hunting proximity speeds $\Delta v$. (b) Recursive %AutoRegressive (RAR) model-based `frozen' Time Varying Power Spectral Density (TV-PSD) %estimate at hunting onset ($\Delta v=0$).  [$f_h$ designates the hunting frequency; RAR %modeling details: AR order $n=8$, forgetting factor $\lambda=0.9995$.]} 
%\label{fig:Indicative_Signals}
%\end{center}
%\end{figure}

\section{The postulated hunting proximity assessment method}\label{S.4}

\subsection{The development of a grey vehicle model}\label{S.4.1}

The Jacobian matrix $\boldsymbol{J}(\boldsymbol{x})$ required for MLE calculation is typically unknown and, as already noted, its estimation using state-of-the-art methods is extremely challenging for high-order, complex systems \citep{PARLITZ2016}, such as railway vehicles. To address this challenge, a \emph{grey}, \emph{nonlinear} vehicle model is developed, featuring a physics-informed structure with unknown parameters estimated from acceleration measurements.
The grey model is obtained from the simulation model of \autoref{eq:1}, which captures the essential lateral, hunting-related vehicle dynamics, through polynomial approximation of the nonlinear and non-smooth wheel–rail contact parameters in terms of $y_{ei}$ ($i=1,\ldots,4$). This ensures accurate representation of the contact nonlinearities without unnecessary model complexity, while motions and terms with negligible influence on the lateral dynamics, such as the car-body roll angle $\phi_c$, are deliberately omitted.
As a result, a more parsimonious grey model, with fewer and identifiable parameters than the simulation model (described by \autoref{eq:1}), is formulated as follows:
\begin{equation}\label{eq:Grey_Model}
\begin{aligned}
    {}& \Ddot{y}_{w1} = [y_{w1}\;\dot{y}_{w1}\;\psi_{w1}\;\dot{\psi}_{w1}\;y_{b1}\;\dot{y}_{b1}\;\psi_{b1}\;\dot{\psi}_{b1}\;\phi_{b1}\;\dot{\phi}_{b1}\;y_{e1}\:y_{e1}^3\;\dot{y}_{w1}\!\cdot y_{e1}^2\;\psi_{w1}\!\cdot y_{e1}^2] \cdot [a_{1,1}\ldots a_{1,14}]^T \\
    {}& \Ddot{\psi}_{w1} = [\psi_{w1}\;\dot{\psi}_{w1}\;\psi_{b1}\;\dot{\psi}_{b1}\;y_{e1}\;y_{e1}^3\;y_{e1}^5\;y_{e1}^7\:\dot{\psi}_{w1}\!\cdot y_{e1}^2] \cdot [a_{2,1}\ldots a_{2,9}]^T \\
    {}& \Ddot{y}_{w2} = [y_{w2}\;\dot{y}_{w2}\;\psi_{w2}\;\dot{\psi}_{w2}\;y_{b1}\;\dot{y}_{b1}\;\psi_{b1}\;\dot{\psi}_{b1}\;\phi_{b1}\;\dot{\phi}_{b1}\;y_{e2}\;y_{e2}^3\;\dot{y}_{w2}\!\cdot y_{e2}^2\;\psi_{w2}\!\cdot y_{e2}^2] \cdot [a_{3,1}\ldots a_{3,14}]^T \\
    {}& \Ddot{\psi}_{w2} = \big[\psi_{w2}\;\dot{\psi}_{w2}\;\psi_{b1}\;\dot{\psi}_{b1}\;y_{e2}\;y_{e2}^3\;y_{e2}^5\;y_{e2}^7\;\dot{\psi}_{w2}\!\cdot y_{e2}^2\big]  \cdot [a_{4,1} \ldots a_{4,9}]^T \\
   {}& \Ddot{y}_{w3} = [y_{w3}\;\dot{y}_{w3}\;\psi_{w3}\;\dot{\psi}_{w3}\;y_{b2}\;\dot{y}_{b2}\;\psi_{b2}\;\dot{\psi}_{b2}\;\phi_{b2}\;\dot{\phi}_{b2}\;y_{e3}\;y_{e3}^3\;\dot{y}_{w3}\!\cdot y_{e3}^2\;\psi_{w3}\!\cdot y_{e3}^2] \cdot [a_{5,1}\ldots a_{5,14}]^T\\
     {}& \Ddot{\psi}_{w3} = [\psi_{w3}\;\dot{\psi}_{w3}\;\psi_{b2}\;\dot{\psi}_{b2}\;y_{e3}\;y_{e3}^3\;y_{e3}^5\;y_{e7}^3 \;\dot{\psi}_{w3}\!\cdot y_{e3}^2] \cdot [a_{6,1}\ldots a_{6,9}]^T\\
%{}&\text{\emph{Rear wheelset of the trailing bogie}}\!: \\[-1ex]
     {}& \Ddot{y}_{w4} = [y_{w4}\;\dot{y}_{w4}\;\psi_{w4}\;\dot{\psi}_{w4}\;y_{b2}\;\dot{y}_{b2}\;\psi_{b2}\;\dot{\psi}_{b2}\;\phi_{b2}\;\dot{\phi}_{b2}\;y_{e4}\;y_{e4}^3\;\dot{y}_{w4}\!\cdot y_{e4}^2\;\psi_{w4}\!\cdot y_{e4}^2] \cdot [a_{7,1} \ldots a_{7,14}]^T \\
     {}& \Ddot{\psi}_{w4} = [\psi_{w4}\;\dot{\psi}_{w4}\;\psi_{b2}\;\dot{\psi}_{b2}\;y_{e4}\;y_{e4}^3\;y_{e4}^5\;y_{e4}^7\;\dot{\psi}_{w4}\!\cdot y_{e4}^2] \cdot [a_{8,1} \ldots a_{8,9}]^T\\
     {}& \Ddot{y}_{b1} = [y_{w1}\;\dot{y}_{w1}\;y_{w2}\;\dot{y}_{w2}\;y_{b1}\;\dot{y}_{b1}\;\psi_{b1}\;\dot{\psi}_{b1}\;\phi_{b1}\;\dot{\phi}_{b1}\;y_c\;\dot{y}_c\;\psi_{c}\;\dot{\psi}_{c}] \cdot [a_{9,1} \ldots a_{9,14}]^T\\
     {}& \Ddot{\psi}_{b1} = [y_{w1}\;\dot{y}_{w1}\;\psi_{w1}\;\dot{\psi}_{w1}\;y_{w2}\;\dot{y}_{w2}\;\psi_{w2}\;\dot{\psi}_{w2}\;y_{b1}\;\dot{y}_{b1}\;\psi_{b1}\;\dot{\psi}_{b1}\;\phi_{b1}\;\dot{\phi}_{b1}\;\psi_{c}\;\dot{\psi}_{c}] \cdot [a_{10,1} \ldots a_{10,16}]^T \\
     {}& \Ddot{\phi}_{b1} =[y_{w1}\;\dot{y}_{w1}\;y_{w2}\;\dot{y}_{w2}\;y_{b1}\;\dot{y}_{b1}\;\psi_{b1}\;\dot{\psi}_{b1}\;\phi_{b1}\;\dot{\phi}_{b1}\;y_{c}\;\dot{y}_{c}\;\psi_{c}\;\dot{\psi}_{c}\;y_{e1}\;y_{e2}\;y_{e1}^3\;y_{e2}^3] \cdot [a_{11,1} \ldots a_{11,18}]^T\\
    {}& \Ddot{y}_{b2} = [y_{w3}\;\dot{y}_{w3}\;y_{w4}\;\dot{y}_{w4}\;y_{b2}\;\dot{y}_{b2}\;\psi_{b2}\;\dot{\psi}_{b2}\;\phi_{b2}\;\dot{\phi}_{b2}\;y_c\;\dot{y}_c\;\psi_{c}\;\dot{\psi}_{c}]  \cdot  [a_{12,1} \ldots a_{12,14}]^T \\
    {}& \Ddot{\psi}_{b2} = [y_{w3}\;\dot{y}_{w3}\;\psi_{w3}\;\dot{\psi}_{w3}\;y_{w4}\;\dot{y}_{w4}\;\psi_{w4}\;\dot{\psi}_{w4}\;y_{b2}\;\dot{y}_{b2}\;\psi_{b2}\;\dot{\psi}_{b2}\;\phi_{b2}\;\dot{\phi}_{b2}\;\psi_{c}\;\dot{\psi}_{c}] \cdot [a_{13,1} \ldots a_{13,16}]^T \\
     {}& \Ddot{\phi}_{b2} = [y_{w3}\;\dot{y}_{w3}\;y_{w4}\;\dot{y}_{w4}\;y_{b2}\;\dot{y}_{b2}\;\psi_{b2}\;\dot{\psi}_{b2}\;\phi_{b2}\;\dot{\phi}_{b2}\;y_{c}\;\dot{y}_{c}\;\psi_{c}\;\dot{\psi}_{c}\;y_{e3}\;y_{e4}\;y_{e3}^3\;y_{e4}^3] \cdot [a_{14,1} \ldots a_{14,18}]^T \\
    {}& \Ddot{y}_{c} =\big[y_{b1}\;\dot{y}_{b1}\;\phi_{b1}\;\dot{\phi}_{b1}\;y_{b2}\;\dot{y}_{b2}\;\phi_{b2}\;\dot{\phi}_{b2}\;y_{c}\;\dot{y}_{c}\;\psi_{c}\;\dot{\psi}_{c}\big] \cdot [a_{15,1} \ldots a_{15,12}]^T\\
     {}& \Ddot{\psi}_{c} = \big[y_{b1}\;\dot{y}_{b1}\;\psi_{b1}\;\dot{\psi}_{b1}\;\phi_{b1}\;\dot{\phi}_{b1}\;y_{b2}\;\dot{y}_{b2}\;\psi_{b2}\;\dot{\psi}_{b2}\;\phi_{b2}\;\dot{\phi}_{b2}\;y_{c}\;\dot{y}_{c}\;\psi_{c}\;\dot{\psi}_{c}\big] \cdot [a_{16,1} \ldots a_{16,16}]^T
\end{aligned} 
\end{equation}
with the model being valid for wheelset excursions $y_{ei}$ ($i = 1,\ldots,4$) not exceeding the flange clearance $\eta$. Each grey-model equation is expressed in linear-in-parameters form and associated with parameter vector $\boldsymbol{\theta}_l:=[a_{l,1}\; a_{l,2} \ldots a_{l,n_l}]^T$, where $n_l$ designates the number of parameters in the $l$-th equation ($l=1,\ldots,16$). The complete (unified) grey-model parameter vector is then defined by stacking these individual vectors as $\boldsymbol{\vartheta}:=[\boldsymbol{\theta}_{1}^T\;\boldsymbol{\theta}_{2}^T\ldots \boldsymbol{\theta}_{16}^T]^T$. Assuming a constant vehicle speed $v$, the unified parameter vector $\boldsymbol{\vartheta}$ is time-invariant and can therefore be estimated via Linear Least Squares (LLS) \citep[p. 206]{LJUNG1999} using a vector acceleration signal $\bm{y}[k]$, whose scalar components $y_j[k]$ ($j=1,\ldots,22)$, are measured by 22 uni-axial sensors ($s=22$) mounted on the railway vehicle at the locations shown in \autoref{fig:Simulation_Model_A}. The estimation of a continuous-time model from discrete-time measurements is valid for sufficiently small sampling periods $T_s$ (typically $T_s<0.1/f_c$, where $f_c$ designates the signal bandwidth in Hz \citep{RAO2006}). For further details on grey-model parameter estimation and the signals employed, the reader is referred to \autoref{App:B} and \autoref{Signal_Derivation}, respectively.

\begin{table}[!t]
\centering
	\caption{Details on the deduction of the required signals for grey model parameter estimation.}	
    {
\renewcommand{\arraystretch}{1.0}
{\begin{tabular}{cl}
\hline
Component & \multicolumn{1}{c}{Signal(s)} \\
\hline
$i$-th wheelset\textsuperscript{(a)} & $\Ddot{y}_{wi}=y_{3i-1},\; \Ddot{\psi}_{wi}=\tan^{-1}\Big(\dfrac{y_{3i}-y_{3i-2}}{2d_w}\Big),\;\Ddot{x}_{wi}=\dfrac{y_{3i-2}+y_{3i}}{2}$   \\ 
$j$-th bogie\textsuperscript{(a)} & $\Ddot{y}_{bj}=y_{4j+11},\; \Ddot{\psi}_{bj}=\tan^{-1}\Big(\dfrac{y_{4j+9}-y_{4j+11}}{2d_w}\Big),\;\Ddot{\phi}_{bj}=\tan^{-1}\Big(\dfrac{y_{4j+12}-y_{4j+10}}{l_c}\Big)$   \\
car-body & $\Ddot{y}_{c}=y_{22},\; \Ddot{\psi}_{c}=\tan^{-1}\Big(\dfrac{y_{21}-y_{22}}{l_c}\Big)$   \\
\hline
 \multicolumn{2}{l}{\small{The velocity ($\dot{y}_{wi},\:\dot{\psi}_{wi},\:\dot{x}_{wi}\:\dot{y}_{bj},\:\dot{\psi}_{bj},\:\dot{\phi}_{bj},\:\dot{y}_{c},\:\dot{\psi}_{c}$) and displacement ($y_{wi}$, $\psi_{wi}$, $x_{wi}$, $y_{bj}$, $\psi_{bj}$, $\phi_{bj}$, $y_{c}$, $\psi_{c}$)}}\\
\multicolumn{2}{l}{\small{signals are obtained via numerical integration using the trapezoidal rule \citep{WORDEN1990}.}}\\ \hline
\multicolumn{2}{l}{\small{\textsuperscript{(a)}$i=1,\ldots,4$; $j=1,2$.}}
\end{tabular}}}

\label{Signal_Derivation}
\end{table}

In practice, the wheelset excursion $y_{ei}$ ($i=1,\ldots,4)$ is not available as the track irregularity excitation $u_{i}$ is generally unknown. By combining the equations of motion in \autoref{eq:1} for the longitudinal deviation $x_{wi}$ and yaw angle $\psi_{wi}$ of wheelset $i$ and neglecting insignificant terms the following expressions are obtained:
\begin{equation}\label{eq:Fx_Calc}
\Ddot{x}_{wi} =  \frac{2}{m_w}F_{xi}^{(l,r)} \mp \frac{I_{wz}}{m_wa_0}\Ddot{\psi}_{wi} - \frac{2k_{pxi}}{m_w}\Big[x_{wi} \mp \frac{d_w^2}{a_0}(\psi_{wi} -\psi_{bj})\Big] - \frac{2c_{pxi}}{m_w}\Big[\dot{x}_{wi}\mp \frac{d_w^2}{a_0}(\dot{\psi}_{wi}-\dot{\psi}_{bj})\Big]
\end{equation}
in which the upper and lower signs of $\mp$ correspond to the left and right wheels, respectively. 
By neglecting terms with minor influence and the saturation coefficients $\varepsilon_i$, and approximating the creep forces $F_{xi}^{(l,r)}$ as linear functions of $y_{wi}$ and $u_a$, \autoref{eq:Fx_Calc} is rewritten as: 
\begin{equation}\label{eq:Track_Irreg_Form}
     \Ddot{x}_{wi} =\pm p_1(y_{wi} -u_i) + p_2\dot{\psi}_{wi}u_i \pm p_3\dot{\psi}_{wi} \pm p_4\Ddot{\psi}_{wi} - p_5x_{wi} \pm p_6(\psi_{wi} -\psi_{bj}) - p_7\dot{x}_{wi}\pm {p}_8(\dot{\psi}_{wi} -\dot{\psi}_{bj})
\end{equation}
where the upper and lower signs of $\pm$ correspond to $\dot{\psi}_{wi} > 0$ (left wheels) and $\dot{\psi}_{wi} < 0$ (right wheels), respectively, and $\{{p}_1,\ldots,{p}_8\}$ represents the set of unknown parameters to be determined for estimating $u_i$. Owing to the adopted linear approximations, \autoref{eq:Track_Irreg_Form} becomes less accurate for larger wheelset lateral excursions $y_{ei}$.  All wheels are excited by the same track irregularity profile, with appropriate time shifts. For instance, for the first two wheelsets, $u_1[k]=u_2[k+L]$, where $L$ designates the closest integer to $l_b/(v T_s)$. Combining this relation with \autoref{eq:Track_Irreg_Form} for the left wheels of the first two wheelsets ($i = 1,2$)\footnote{An equivalent formulation can be obtained by considering any other pair of wheelsets or, alternatively, the corresponding right wheels.} eliminates the unknown track irregularity excitation $u_i$ and enables the estimation of the parameters $\{{p}_1,\ldots,{p}_8\}$ via the LLS method. Using these parameter estimates and \autoref{eq:Track_Irreg_Form}, the wheelset excursion $y_{ei}$ is obtained as follows\footnote{The hat designates estimator/estimate.}:
\begin{equation}\label{eq:Irreg}
y_{ei} = y_{wi}-\dfrac{\Ddot{x}_{wi}  \mp {\widehat{{p}}}_1y_{wi} \mp {\widehat{{p}}}_3\dot{\psi}_{wi} \mp {\widehat{{p}}}_4\Ddot{\psi}_{wi}+{\widehat{{p}}}_5x_{wi}\mp{\widehat{{p}}}_6(\psi_{wi}-\psi_{bj})+{\widehat{{p}}}_7\dot{x}_{wi}\pm{\widehat{{p}}}_8\dot{\psi}_{bj}}{ {\widehat{{p}}}_2\dot{\psi}_{wi}\mp{\widehat{{p}}}_1}
\end{equation}
where the upper and lower signs of $\pm$ and $\mp$ correspond to $\dot{\psi}_{wi} > 0$ and $\dot{\psi}_{wi} < 0$, respectively.
The expression of $y_{ei}$ explicitly involves the longitudinal acceleration $\Ddot{x}_{wi}$, velocity $\dot{x}_{wi}$ and displacement $x_{wi}$, which are not state variables of the grey model defined in \autoref{eq:Grey_Model}. Consequently, the model is non-autonomous. An equivalent autonomous state-space representation (refer to the next subsection) is obtained by augmenting the state vector with time $t$ \citep[pp. 2–3]{PARKER1989}. Moreover, due to the piecewise-smooth structure of $y_{ei}$, the Jacobian matrix $\boldsymbol{J}(\boldsymbol{x})$ of the grey model is computed analytically within each active regime. Further details regarding the state-space formulation and Jacobian matrix computation  are provided in \autoref{App:C}.

\subsection{The Maximum Lyapunov Exponent (MLE)}\label{S.4.2}
The postulated railway hunting proximity assessment method is founded upon the concept of the Maximum Lyapunov Exponent (MLE) \cite{ABARBANEL1993,PARLITZ2016}. Consider an autonomous nonlinear continuous-time system in state-space form: 
\begin{equation}\label{eq:Theor_Model_State_Space}
    \boldsymbol{\Dot{x}}={} \boldsymbol{f}(\boldsymbol{x}) 
\end{equation}
with $\boldsymbol{x}$ ($n\times 1$) designating the state vector, $n$ the system order, and $\boldsymbol{f}(\boldsymbol{x})={}[f_1(\boldsymbol{x}),\ldots, f_n(\boldsymbol{x})]^{T}$ a set of nonlinear functions of  $\boldsymbol{x}$ (with $^{T}$ designating transposition). Let $\boldsymbol{x}(t)$ be the solution (trajectory) of \autoref{eq:Theor_Model_State_Space} starting from the initial condition $\boldsymbol{x}(0)$. Applying a small perturbation $\delta{\boldsymbol{x}}(0)$ to the initial condition yields the perturbed trajectory $\boldsymbol{x}'(t)=\boldsymbol{x}(t)+\delta{\boldsymbol{x}}(t)$. Assuming that $\boldsymbol{f}(\boldsymbol{x})$ is continuously differentiable in a neighborhood of $\boldsymbol{x}(t)$, linearization of \autoref{eq:Theor_Model_State_Space} along the solution $\boldsymbol{x}(t)$ yields the evolution of the distance between the two trajectories $\delta\boldsymbol{x}(t)$ as follows \cite[p. 75]{PARKER1989}:
\begin{equation}\label{eq:Variational_Equations}
    \delta \boldsymbol{\Dot{x}}={} \boldsymbol{J}(\boldsymbol{x}(t))\delta \boldsymbol{x}
\end{equation}
with $\boldsymbol{J}(\boldsymbol{x}(t))=(\partial\boldsymbol{f}/\partial\boldsymbol{x})_{\boldsymbol{x}=\boldsymbol{x}(t)}$ designating the system's Jacobian matrix. Let $\delta{\boldsymbol{x}(t)}$ be the solution of \autoref{eq:Variational_Equations} starting from an arbitrarily selected initial distance $\delta{\boldsymbol{x}}(0)$. For large enough $t$, the MLE of the system of \autoref{eq:Theor_Model_State_Space} can be obtained as \citep[p. 75]{PARKER1989}:
\begin{equation}\label{eq:MLE}
\text{MLE}= \frac{1}{t} \ln\bigg(\frac{\| \delta{\boldsymbol{x}(t)}\|_2}{\| \delta{\boldsymbol{x}(0)}\|_2}\bigg)
\end{equation}
with $\|\cdot\|_2$ designating Euclidean norm. In the limit as $t\to \infty$, the MLE converges to its true value. In practice, the $\delta{\boldsymbol{x}(t)}$ is typically computed via numerical integration \citep[pp. 73-74]{PARKER1989}. However, for sufficiently small time intervals, the Jacobian may be assumed locally constant, allowing  $\delta{\boldsymbol{x}(t)}$ to be obtained analytically as the solution of a first-order linear ODE over each interval.

\subsection{A brief summary of the proposed method}\label{S.4.3}

The postulated method operates under four assumptions:
(i) approximately constant vehicle speed $v$, (ii) sufficiently long measured signals, (iii) accurate representation of the vehicle’s lateral (hunting-related) dynamics by the estimated grey model, and (iv) operation initiated while the train is at rest, with zero initial conditions.
Assumptions (i) and (ii) ensure convergence of the computed MLE to its theoretical true value (see \autoref{eq:MLE}), while assumption (iii) ensures that the Jacobian matrix $\boldsymbol{J}(\boldsymbol{x})$ closely approximates that of the actual vehicle, enabling accurate MLE estimation (see \autoref{eq:Variational_Equations}). Assumption (iv) provides the known initial conditions required for numerical integration, thereby preventing drift in the derived signals.
In summary, the \emph{Operational Phase} of the proposed method consists of the following steps: \vspace*{2mm} \\
\noindent \textbf{Step 1: Signal acquisition and track irregularity estimation.}  \underline{Substep 1.1:} As a first step, a vector acceleration signal $\bm{y}[k]$ is acquired using 22 uni-axial on-board sensors on locations as shown in \autoref{fig:Simulation_Model_A}. The corresponding velocity and displacement signals are subsequently obtained via numerical integration using the trapezoidal rule \citep{WORDEN1990} (see \autoref{Signal_Derivation} for details). The vehicle speed $v$ is measured from an onboard tachometer, while the wheelbase length $2l_b$ is an easily available vehicle design parameter.  \underline{Substep 1.2:} The specific track irregularity profiles $u_i$ ($i=1,\ldots,4$), and thus the wheelset excursion signals $y_{ei}$, are obtained using \autoref{eq:Irreg}, with the associated parameters $\{{p}_1,\ldots,{p}_8\}$ estimated via the LLS algorithm. \vspace*{-1mm}

\noindent \textbf{Step 2: Grey model estimation and Jacobian matrix computation.} \underline{Substep 2.1:} Using the signals obtained in \emph{Step 1}, the complete grey model parameter vector $\boldsymbol{\vartheta}$ is estimated via the LLS algorithm (see details in \autoref{App:B}). \underline{Substep 2.2:} Based on the estimated grey model parameter vector $\boldsymbol{\vartheta}$ and the signals obtained in \emph{Step 1}, the Jacobian matrix of the grey model is computed analytically and evaluated for $k = 1,2,\ldots,N$, yielding simply $\boldsymbol{J}[k]$  (see details in \autoref{App:C}). \vspace*{-1mm}
    
\noindent  \textbf{Step 3: Maximum Lyapunov Exponent (MLE) estimation.} \underline{Substep 3.1:} Although the MLE in \autoref{eq:MLE} is defined in the limit as $t\to \infty$, the solution $\boldsymbol{\delta x}(t)$ of \autoref{eq:Variational_Equations} grows or decays exponentially with time, rendering its norm numerically unreliable for large $t$ \cite[pp. 73-74]{PARKER1989}. To preserve numerical stability, the sampled version $\boldsymbol{\delta x}[k]$ of $\boldsymbol{\delta x}(t)$ is normalized every $M$ samples at time instants $k_i=iM$ ($i=1,2,\ldots,\lfloor N/M\rfloor$; $k_{0}=0$), hereafter referred to as MLE estimation time instances. The parameter $M$ is user-defined, and $\lfloor \cdot \rfloor$ denotes the floor operator. Let $\delta{\boldsymbol{\bar{x}}}[k_{i}]$ denote $\delta{\boldsymbol{x}}[k_{i}]$ normalized by its Euclidean norm, and set $\delta{\boldsymbol{\bar x}}[0]=[1\;0 \ldots 0]^T$. \underline{Substep 3.2:} Starting from $\delta{\boldsymbol{\bar x}}[k_{i-1}]$ and using $\boldsymbol{J}[k]$ obtained in \emph{substep 3.1}, the analytical solution of \autoref{eq:Variational_Equations} (assuming $\boldsymbol{J}[k]$ constant over each sampling interval) is applied sequentially as: 
    \begin{equation}\label{eq:Discrete_Var_Sol}
        \delta{\boldsymbol{x}}[k+1]= e^{\boldsymbol{J}[k]T_s}\delta{\boldsymbol{x}}[k],\quad k=k_{i-1},\ldots,k_i-1,\quad \delta{\boldsymbol{x}}[k_{i-1}]:=\delta{\boldsymbol{\bar x}}[k_{i-1}] 
    \end{equation}
    so that $\delta{\boldsymbol{x}}[k_i]$ is obtained. Based on \autoref{eq:MLE}, the MLE is estimated at $k_i$ as \citep[p. 75]{PARKER1989}: 
\begin{equation}\label{eq:MLE_Final}
\text{MLE}[k_i]= \frac{1}{k_iT_s} \sum_{j=1}^{i} \ln(\| \delta{\boldsymbol{x}[k_j]}\|_2)
\end{equation} 
\underline{Substep 3.3:}  Normalize $\delta{\boldsymbol{x}}[k_{i}]$ by its Euclidean norm to obtain $\delta{\boldsymbol{\bar{x}}}[k_{i}]$, increase $i$ by one, and go back to \emph{substep 3.2}, as long as $i\leq \lfloor N/M\rfloor$.  \vspace*{-1mm}

\noindent \textbf{Step 4: Hunting proximity assessment.} The final estimated value MLE[$k_{\lfloor N/M\rfloor}$] is classified according to user-defined threshold ranges, each of which corresponds to a specific considered proximity interval $\Delta v_{i}$ ($i=1,\ldots,8$; see for details \hyperref[eq:Prox_Intervals]{Equation~\ref*{eq:Prox_Intervals}}), thus assessing the proximity to hunting onset.

 The design parameters $M$, $N$ and $T_s$ are determined during an initial \emph{Tuning Phase} using a single simulated or physical experiment under approximately constant speed $v$.  The normalization step $M$ is selected according to \citep[p. 81]{PARKER1989}.
The signal length $N$ is selected to ensure convergence of the MLE estimate, as described in \citep{BALCERZAK2020}, while the sampling period $T_s$ (in s) is selected to be sufficiently small ($T_s<0.1/f_c$) to ensure accurate estimation of the continuous-time grey model \citep{RAO2006}. 

\section{Performance assessment via Monte Carlo simulations}\label{S.5}

The design parameters are selected following \autoref{S.4.3} using a single Monte Carlo simulation under nominal vehicle/track conditions (scenario A; $\Delta v \in \Delta v_8$); the selected values are summarized in \autoref{Design_Parameters}.

%---------------------
\begin{table}[!b]
    \centering
	\caption{Selected design parameters.}	
	{\begin{tabular}{>{\centering\arraybackslash}p{4.5cm}
                >{\centering\arraybackslash}p{4cm}
                >{\centering\arraybackslash}p{4cm}}
			\hline
			Normalization Step ($M$)\textsuperscript{(a)} & Signal Length ($N$) & Sampling Period ($T_s$)  \\
            \hline
			$5\:000$ samples [0.25 s] &  $1.6\times 10^{6}$ samples [$80\ $s] &  $5\times 10^{-5}\ $s \\
			\hline
            \multicolumn{3}{l}{\small{\textsuperscript{(a)}Excitation frequency range: $f_c=v_{\max}/L_{\min}\approx 40$ Hz, where $L_{\min}=3$ m (track irregularity}}\\
            \multicolumn{3}{l}{\small{lower wavelength limit) and $v_{\max}=118.61$ m/s (maximum operating speed).}}\\
	\end{tabular}}
	\label{Design_Parameters}
\end{table}

\begin{figure}[!t]
\begin{center}
\includegraphics[width=\textwidth,trim={1.1cm 0.425cm 2.6cm 0.1cm},clip]{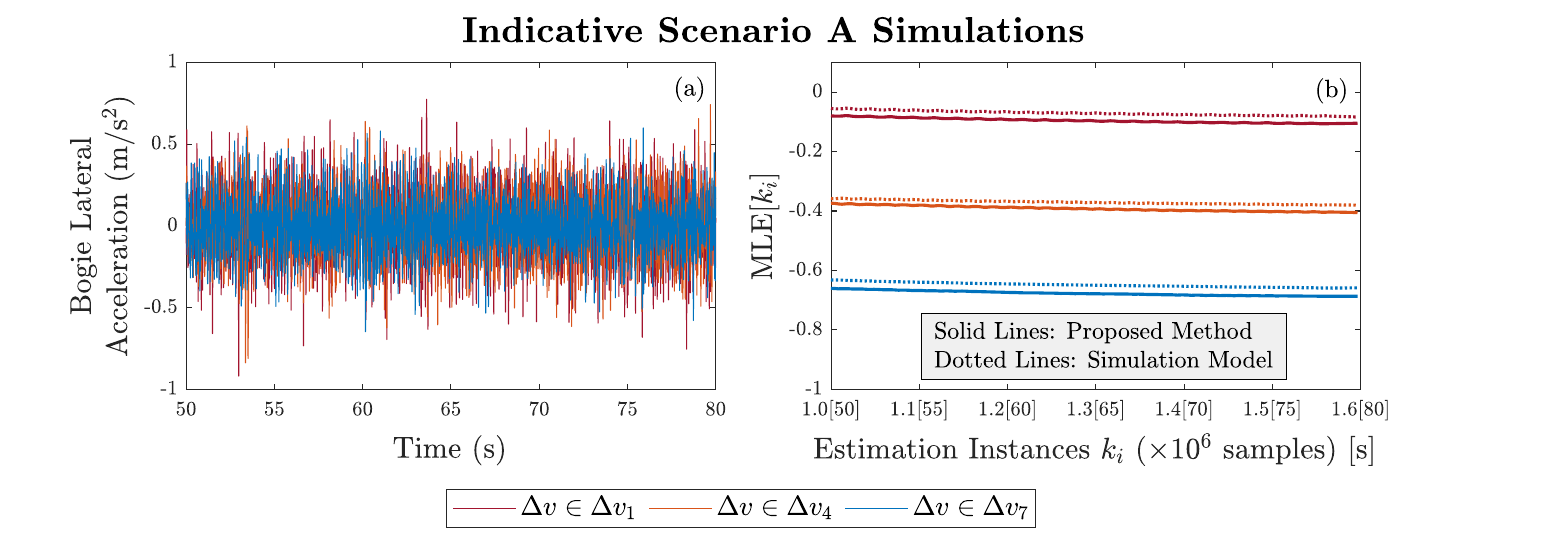} 
\caption{Indicative scenario A Monte Carlo simulations for different hunting proximity intervals: (a) Bogie lateral acceleration signals (sensor 15). (b) MLE estimates obtained with the proposed method (solid lines) and the simulation model (dotted lines).} 
\label{fig:Indicative_Signals_MLE}
\end{center}
\end{figure}

Performance/robustness assessment is carried out within an \emph{Operational Phase} based on 15 Monte Carlo simulations per considered operating scenario and hunting proximity interval (480 in total; for details refer to \autoref{Table:Monte_Carlo_Sims}).
The MLE estimates for representative simulations under operating scenario A (nominal track/vehicle) and selected hunting proximity intervals are presented in \autoref{fig:Indicative_Signals_MLE}. For comparison, the true MLE estimates are also shown (dotted lines) which are computed according to Step 3 of the proposed method (see \autoref{S.4.3}), using the Jacobian matrix $\boldsymbol{J}(\boldsymbol{x})$ of the simulation model\footnote{Since closed-form (analytical) expressions for $\boldsymbol{f}(\boldsymbol{x})$ are not available - owing to the wheel–rail contact relations being implemented via look--up tables (refer to  \autoref{S.3.1}) -- $\boldsymbol{J}(\boldsymbol{x})$ is obtained numerically using Newton’s forward finite--difference scheme. At points of discontinuity the Jacobian $\boldsymbol{J}(\boldsymbol{x})$ is not evaluated, and the last valid value is retained for MLE estimation.}.
It is evident that the proposed method yields MLE estimates that are in close agreement with the true values for all simulations. Furthermore, the MLE exhibits a consistent evolution, increasing as the vehicle speed $v$ approaches the hunting speed $v_h$, and consequently as the proximity measure $\Delta v$ decreases ($\Delta v_1<\Delta v_4<\Delta v_7$; see \hyperref[eq:Prox_Intervals]{Equation~\ref*{eq:Prox_Intervals}}).

\begin{figure}[!t]
\begin{center}
\includegraphics[width=0.985\textwidth,trim={0.55cm 0.4cm 3.1cm 0cm},clip]{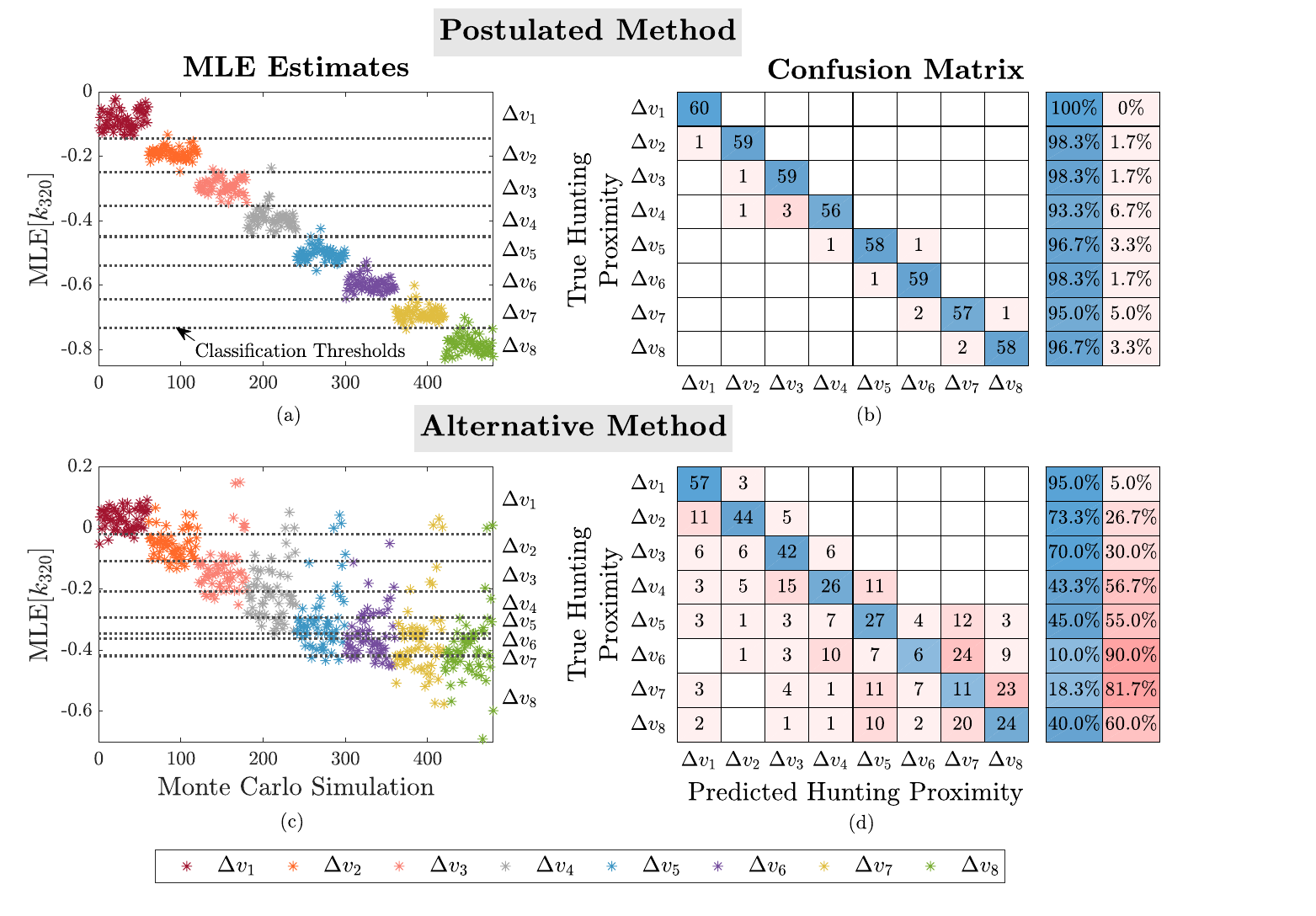} 
\caption{Aggregate performance of the postulated and alternative methods over all operating scenarios, proximity intervals, and Monte Carlo simulations: (a,c) final MLE estimates, MLE[$k_{320}$], and (b,d) corresponding confusion matrices. [120 Monte Carlo simulations per scenario; 480 simulations in total.]} 
\label{fig:Perf_Results}
\end{center}
\end{figure}

Cumulative hunting proximity assessment results, encompassing all operating scenarios and Monte Carlo simulations, are presented in \hyperref[fig:Perf_Results]{Figure~\ref{fig:Perf_Results}(b)} through a confusion matrix. In this, each row $i$ corresponds to a true hunting proximity interval $\Delta v_i$ and each column $j$ to a predicted interval $\Delta v_j$, with $i,j\in\{1,\ldots,8\}$.  The entry ($i,j$) represents the number of simulations with true interval $\Delta v_i$ classified (predicted) as $\Delta v_j$; thus, thus, correct and incorrect classifications appear on the diagonal and off-diagonal entries, respectively, with their percentages reported in the two rightmost columns. Classification decisions are based on the final MLE estimates (MLE[$k_{320}$]; with $k_{320}=1.6\times 10^{6}$ samples), using optimal thresholds that maximize ACC. The corresponding MLE values and thresholds are presented in \hyperref[fig:Perf_Results]{Figure~\ref{fig:Perf_Results}(a)}.
Evidently, the proposed hunting proximity assessment method attains excellent performance, achieving an overall ACC of $97.08\%$. It is worth emphasizing that misclassifications are limited to neighboring proximity intervals, indicating no substantial under- or over-estimation of hunting proximity.

The method's performance robustness to uncertainty arising from suspension faults, worn wheels and worn tracks is presented in \autoref{Robustness} in terms of ACC metric. Optimal classification thresholds and the final MLE estimates are employed (see \hyperref[fig:Perf_Results]{Figure~\ref{fig:Perf_Results}(a)}). For details on the scenarios, see \autoref{Operating_Scenarios}.
As is evident, the proposed method achieves excellent performance ($\text{ACC}\geq99.17\%$) for scenarios A$-$C, while the ACC decreases to $89.17\%$ for scenario D; however the performance is still considered satisfactory.  Worn track conditions (scenario D) increase wheelset lateral deviations toward the flange clearance, thereby reducing both the grey model’s ability to represent the actual vehicle dynamics -- here, the simulation model dynamics -- and the accuracy of track-irregularity estimation (see \autoref{S.4.1}). The method demonstrates excellent overall robustness, which is somewhat compromised under worn track conditions.

%---------------------
\begin{table}[!t]
\centering
	\caption{Performance robustness assessment under each operating scenario in terms of ACC (\%).}	
	{\begin{tabular}{c c c c c c}
			\hline
            Operating Scenario(s) & A & B & C & D & All \\ 			\hline
             ACC &  100\% & 100\% & 99.17\% & 89.17\% & 97.08\%\\
           \hline
	\end{tabular}}
	\label{Robustness}
\end{table}

The proposed method is directly compared with an indirect (data-driven) MLE estimation method reported in \cite{ESCOT2023}, suitably adapted to the present framework. In this method, the unknown nonlinear dynamics are first identified from signal measurements via the use of local polynomial kernel regression. The Jacobian matrix $\boldsymbol{J}(\boldsymbol{x})$ of the resulting model is then derived analytically and used to compute the MLE following \emph{Step 3} of our method, with the resulting MLE used for hunting proximity assessment following \emph{Step 4} (see \autoref{S.4.3}). To ensure a fair comparison, all displacement and velocity signals obtained in \emph{substep 1.1} of the proposed method (see \autoref{S.4.3}) are directly used to construct the state-space trajectory $\boldsymbol{x}[k]$, as in \autoref{App:C}. Since the present problem includes exogenous inputs, the method is accordingly adapted by augmenting the regression model inputs with the track irregularity profiles 
$u_i$ ($i=1,\ldots,4$), as estimated in \emph{substep 1.2} of our method. The method design parameters, the polynomial order ${q}$ and kernel bandwidth ${h}$ are selected following \cite{ESCOT2023}: ${q}=3$ (without cross-terms), ${h}=2.4$. The remaining parameters associated with MLE estimation ($M,N,T_s$) are selected as in the proposed method (see \autoref{Design_Parameters}).

Cumulative hunting proximity assessment results obtained using the alternative method, encompassing all operating scenarios and Monte Carlo simulations, are presented in \hyperref[fig:Perf_Results]{Figure~\ref{fig:Perf_Results}(d)} in terms of a confusion matrix. In line with our method, the classification decisions are based on the final MLE estimates and optimal thresholds (see \hyperref[fig:Perf_Results]{Figure~\ref{fig:Perf_Results}(c)}). 
Evidently, the alternative method based on state-of-the-art MLE estimation attains poor performance, achieving an ACC of $49.4\%$. These results are consistent with the limited MLE estimation accuracy reported in the relevant literature for high-order, complex systems and further demonstrate the effectiveness of the proposed grey-modeling-based method, which accurately captures the nonlinear lateral (hunting-related) dynamics and enables reliable MLE-based hunting proximity assessment.

%-------------------------------------------------------------
\section{Concluding remarks}\label{S.6}
The feasibility of achieving effective and robust on-board hunting proximity assessment in railway vehicles under any operating conditions has been investigated for the first time by introducing and thoroughly assessing a novel method based on: (i) a physics-informed grey nonlinear railway vehicle model with unknown parameters estimated from on-board acceleration measurements; and (ii) grey-model-based estimation of the MLE as a measure of proximity to hunting onset.
The performance of the postulated method has been assessed via numerous Monte Carlo simulations under various realistic operating scenarios using a multi-body vehicle model developed in MATLAB. Performance assessment evaluation is based on confusion matrices, as well as, on the Accuracy (ACC) metric.

The main conclusions drawn from the study may be summarized as follows: \vspace*{-1mm} 
\begin{itemize}
    \item[1.] The feasibility of robust on-board hunting proximity assessment is demonstrated through the introduction of a novel method founded upon a physics-informed grey nonlinear railway vehicle model and the MLE as an indicator of proximity to hunting onset. \vspace*{-2mm} 
    \item[2.] The proposed method achieves an ACC of $97.1\%$ in correctly classifying the proximity to hunting onset, $\Delta v$, into eight user-defined proximity intervals $\Delta v_i$ ($i=1,\ldots,8$). \vspace*{-2mm} 
    \item[3.] The postulated method achieves excellent robustness to uncertainty arising from faulty suspension components, and hollow worn wheels; however, its robustness is somewhat compromised for worn track conditions.  \vspace*{-2mm} 
    \item[4.] The method significantly outperforms a properly adapted alternative method, which achieves an ACC of $49.4\%$, highlighting the limitations of existing MLE estimation methods and the superiority of our approach. \vspace*{-1mm} 
\end{itemize}

Evidently, the present study is exploratory and concept-proving. Further work is required for experimental assessment and further refinements.

%%%%%%%%%%%%%%%%%%%%
% Acknowledgements %
%%%%%%%%%%%%%%%%%%%%

\section*{Funding}
This study has been partially funded by the Hellenic Foundation for Research and Innovation (HFRI) under the 3rd Call for HFRI PhD Fellowships (Fellowship No. 6803).

%%%%%%%%%%%%%%
% References %
%%%%%%%%%%%%%%

% Include your references below, e.g. by appending a '.bib' file. 
% The style, IEEE, is already pre-defined in the class description.
\setlength{\bibsep}{-1.5pt}
\bibliography{References}

% If BibTeX is not used, you can enter each bibitem separately using the procedure below, which complies with the IEEE reference style. (When you have less than 10 references, use: "\begin{thebibliography}{1}". Switch to "\begin{thebibliography}{10}" is you have 10 or more references.)

	%\begin{thebibliography}{1}
	%	
	%	\bibitem{heylen1997modal}
	%	W. Heylen, S. Lammens and P. Sas,
	%	\textit{Modal Analysis Theory and Testing}.
	%	Leuven, Belgium: Katholieke Universiteit Leuven, Departement Werktuigkunde, 1997.
	%	
	%	\bibitem{sas1995active}
	%	P. Sas, C. Bao, F. Augusztinovicz and W. Desmet,
	%	``Active control of sound transmission through a double panel partition,''
	%	\textit{Journal of Sound and Vibration}, vol. 180, no. 4, pp. 609-625, 1995.
	%	
	%	\bibitem{boonen2001modified}
	%	R. Boonen and P. Sas,
	%	``Modified smith compensation for feedback active noise control in ducts,''
	%	in \textit{Proceedings of the 2001 International Congress and Exhibition on Noise Control Engineering},
	%   The Hague, The Netherlands, 2001, pp. 619-624.
	%\end{thebibliography}

%%%%%%%%%%%%%%
% Appendices %
%%%%%%%%%%%%%%

% If none, remove the code below.

\newpage
\appendix
\appendixheader

% Reset numbering in Appendix
% Appendix numbering: A.1, B.1, C.1, ...
% Appendix numbering: A.1, B.1, C.1, ...
\numberwithin{equation}{section}
\numberwithin{table}{section}
\numberwithin{figure}{section}

\renewcommand{\theequation}{\thesection.\arabic{equation}}
\renewcommand{\thetable}{\thesection.\arabic{table}}
\renewcommand{\thefigure}{\thesection.\arabic{figure}}

\section{Simulation model parameters \label{App:A}}

\begin{table}[h]
\centering
\setlength{\tabcolsep}{2.5pt}
	\caption{Simulation model parameters.}
	{\begin{tabular}{l l c l}
		\hline
\multicolumn{1}{c}{Parameter(s)} & \multicolumn{1}{c}{Description} & Value(s) & \multicolumn{1}{c}{Unit}\\
\hline		
			$m_c$,$\,m_b$,$\,m_w$   & car-body, bogie frame, and wheelset masses & $17\,781$, $2\,049$,  $2\,132$& kg\\
			$I_{cx}$,$\,I_{cz}$  & car-body roll, and yaw moments of inertia & {$55.95$, $1\,309.95$} & Mg$\cdot$m$^{\mathrm{2}}$\\
			$I_{bx}$,$\,I_{bz}$  & bogie frame roll, and yaw moments of inertia & $1\,314$, $1\,470$, $2\,660$ & Mg$\cdot$m$^{\mathrm{2}}$\\
			$I_{wx}$,$\,I_{wy}$,$\,I_{wz}$  & wheelset roll, pitch, and yaw moments of inertia & $0.47$, $0.03$, $0.47$ & Mg$\cdot$m$^{\mathrm{2}}$\\
			$k_{px}$,$\,k_{py}$,$\,k_{pz}$  & primary stiffness coefficients\textsuperscript{(a)} & $1\,462$, $1\,098$,  $382$& kN/m\\
			$c_{px}$,$\,c_{py}$,$\,c_{pz}$  & primary damping coefficients\textsuperscript{(a)} & $40$, $40$, $20$ & {kN/m$\cdot$s}\\
			{$k_{sx}$,$\,k_{sy}$,$\,k_{sz}$}  & secondary stiffness coefficients\textsuperscript{(a)} & {$110$, $110$, $275$}   & {kN/m}\\
			$c_{sx}$,$\,c_{sy}$,$\,c_{sz}$  & secondary damping coefficients\textsuperscript{(a)}& $20$, 	$20$, $20$ & kN/m$\cdot$s\\
                $k_{ty}$ & flange stiffness coefficient  & 1 700 & MN/m \\ 
                $\eta$ & flange clearance  & 0.0082 & m \\ 
			$l_{c},\,l_{b}$  & longitudinal c.g. distances: car-body/bogie and bogie/wheelset  & $5.75,1.1$ & m\\
			$h_{c},\,h_{b},\,h_{w}$  & vertical distances: car-body c.g./secondary suspension, bogie & $0.72,0.25,0.04$ & m\\
             & c.g./secondary suspension and bogie c.g./primary suspension  &  & \\
			$d_{w}$  & half lateral distance between primary springs/dampers & $1$ & m\\
			$d_{s}$ & half lateral distance between secondary springs/dampers  & $1.008$ & m\\
			$r_{0},\,a_0$  & wheel radius and  half of track gauge& $0.43$, $0.72$ & m\\
            $\mu$  & coefficient of friction  & $0.4$ & \\  
            & wheel/rail profiles  & S1002/UIC60 \\  
            & track irregularity wavelength range & [3--70] & m \\  
			\hline
            \multicolumn{4}{l}{\small{\textsuperscript{(a)}Per axle-box (primary) and per bogie side (secondary) in longitudinal, lateral, and vertical directions, respectively.}}\\
	\end{tabular}}\vspace*{-1.5mm}

	\label{Table_Model_Pars}
\end{table}

\section{Grey model parameter estimation\label{App:B}} 

Since the grey model in \autoref{eq:Grey_Model} is formulated in linear-in-parameters form, and assuming a sufficiently small sampling period $T_s$ \citep{RAO2006}, its $l$-th equation may be written in the following discrete-time regression form as: 
\begin{equation}\label{eq:Discrete_Grey_Model}
    y_l[k]=\boldsymbol{\phi}_l^{T}[k] \boldsymbol{\theta}_l+e_l[k],\qquad e_l[k]\sim \mbox{i.i.d}\; \mathcal{N}(0,\sigma_{e,l}^2),\qquad  l=1,\ldots,16
\end{equation}
with $e_l[k]$ designating the zero-mean innovations (one-step-ahead prediction error) sequence that should be serially uncorrelated and characterized by variance $\sigma_{e,l}^2$, corresponding to factors such as measurement noise and estimation errors, $y_l[k]$ the measured acceleration signal, $\boldsymbol{\phi}_l[k]$ the regression vector, and $\boldsymbol{\theta}_l$ the parameter vector, all associated with the $l$-th grey-model equation.   
Instead of estimating each $\boldsymbol{\theta}_l$ separately, the individual regression models can be stacked to construct a unified vector regression form for estimating the complete grey model parameter vector $\boldsymbol{\vartheta}:=[\boldsymbol{\theta}_{1}^T\;\boldsymbol{\theta}_{2}^T\ldots \boldsymbol{\theta}_{16}^T]^T$ (see also \autoref{S.4.1}). The complete output vector $\boldsymbol{y}[k]$ and the corresponding unified regressor matrix $\boldsymbol{\Phi}[k]$ are defined as:
\begin{equation}
\boldsymbol{\Phi}[k] =
\begin{bmatrix}
\boldsymbol{\phi}_1^{T}[k] & \boldsymbol{0}_{1\times n_2}  & \cdots & \boldsymbol{0}_{1\times n_{16}} \\
\vdots & \vdots & \vdots & \vdots \\
 \boldsymbol{0}_{1\times n_1} & \boldsymbol{0}_{1\times n_2} & \cdots & \boldsymbol{\phi}_{16}^{T}[k]
\end{bmatrix}_{[16\times216]},
\qquad
\boldsymbol{y}[k] =
\begin{bmatrix}
y_{1}[k] \\
\vdots \\
y_{16}[k]
\end{bmatrix}_{[16\times1]}
\end{equation}
where  $\boldsymbol{0}_{1\times n_l}$ represents a zero row vector of length $n_l$, with $n_l$ corresponding to the dimension of the regressor $\boldsymbol{\phi}_l[k]$. Accordingly, the unified grey-model regression equation can be written in compact form as:
\begin{equation}\label{eq:Discrete_Unified}
    \boldsymbol{y}[k]=\boldsymbol{\Phi}[k] \boldsymbol{\vartheta}+\boldsymbol{e}[k],\qquad E\{\boldsymbol{e}[k]\cdot\boldsymbol{e}^T[k]\}=\boldsymbol{\Sigma}_e
\end{equation}
with $\boldsymbol{e}[k]=[e_1[k]\: e_2[k]\ldots e_{16}[k]]^T$ representing the stacked innovations vector, with covariance matrix $\boldsymbol{\Sigma}_e$ that is, in the general case, non-diagonal, thereby accounting for possible cross-correlations among the individual innovations. The complete grey-model parameter vector $\boldsymbol{{\vartheta}}$ can then be obtained using the well-known LLS estimator \citep[p. 206]{LJUNG1999}.

\section{State-space representation and
Jacobian matrix computation\label{App:C}} 

The grey model in \autoref{eq:Grey_Model} is written in autonomous state-space form by appending time $t$ as an additional state variable with unit derivative \citep[pp. 2--3]{PARKER1989}. The resulting state variables yielding the following state-space representation:
\begin{equation}\label{eq:Grey_Model_State_Space}
    \dot{x}_{2j-1} = x_{2j}, \quad \dot{x}_{2j} = f_{j}, \quad \dot{x}_{33} = 1, \quad j=1,\ldots,16
\end{equation}
with $f_j(\boldsymbol{x})$ being generally non-linear functions of the state-space vector $\boldsymbol{x}=[x_1\ldots x_{33}]^T$. For compact representation, each function is written in vector form as $f_{j} (\boldsymbol{x})=\boldsymbol{g}_j(\boldsymbol{x})\cdot \, \widehat{\boldsymbol{\theta}}_j$, where $\widehat{\boldsymbol{\theta}}_j$ represents the estimated individual grey-model parameter vector (see \autoref{App:B}), and $\boldsymbol{g}_j(\boldsymbol{x})$ contains the associated state variables, defined as follows: 
\begin{equation}
\label{eq:phi_compact}
\begin{aligned}
\boldsymbol{g}_1 &= 
\big[ x_1\;\, x_2\;\, x_3\;\, x_4\;\, x_{17}\;\, x_{18}\;\, x_{19}\;\, x_{20}\;\,
x_{21}\;\, x_{22}\;\, y_{e1}\;\, y_{e1}^3\;\, x_1\cdot y_{e1}^2\;\, x_3\cdot y_{e1}^2 \big] \\
\boldsymbol{g}_2 &= 
\big[ x_3\;\, x_4\;\, x_{19}\;\, x_{20}\;\,
y_{e1}\;\, y_{e1}^3\;\, y_{e1}^5\;\, y_{e1}^7\;\, x_4\cdot y_{e1}^2 \big] \\
    {}& \;\; \vdots\\
\boldsymbol{g}_{16} &=
\big[ x_{17}\;\, x_{18}\;\, x_{19}\;\, x_{20}\;\,
x_{21}\;\, x_{22}\;\, x_{23}\;\, x_{24}\;\,
x_{25}\;\, x_{26}\;\, x_{27}\;\, x_{28}\;\,
x_{29}\;\, x_{30}\;\, x_{31}\;\, x_{32} \big]
\end{aligned}
\end{equation}
where the wheelset excursion $y_{ei}$ $(i=1,\dots,4)$ is obtained based on \autoref{eq:Irreg} as follows\footnote{The term $\Ddot{\psi}_{wi}$ is neglected to simplify the resulting state-space model and the computation of $\boldsymbol{J}(\boldsymbol{x})$.}: 
\begin{equation}\label{eq:Irreg_SS}
\left\{ 
\begin{aligned}
     y_{e1}{}&= x_{1}-\dfrac{\Ddot{x}_{w1} \mp \widehat{{p}}_1x_{1} \mp \widehat{{p}}_3x_{4}+\widehat{{p}}_5x_{w1}\mp\widehat{{p}}_6(x_3-x_{19})+\widehat{{p}}_7\dot{x}_{w1} \pm  \widehat{{p}}_8x_{20}}{ \widehat{{p}}_2x_{4}\mp\widehat{{p}}_1} \\ 
    {}& \;\; \vdots\\
    y_{e4}{}&= x_{13}-\dfrac{\Ddot{x}_{w4} \mp \widehat{{p}}_1x_{13} \mp \widehat{{p}}_3x_{16}+\widehat{{p}}_5x_{w4}\mp\widehat{{p}}_6(x_{15}-x_{25})+\widehat{{p}}_7\dot{x}_{w4} \pm  \widehat{{p}}_8x_{26}}{ \widehat{{p}}_2x_{16}\mp\widehat{{p}}_1} \\
\end{aligned}
\right.
\end{equation}
where the upper and lower signs of $\pm$ and $\mp$ correspond to $x_{4i} > 0$ and $x_{4i} < 0$, respectively.

The Jacobian matrix $\boldsymbol{J}(\boldsymbol{x}$) of the state-space model of \autoref{eq:Grey_Model_State_Space} is defined as: 
\begin{equation}\label{eq:Matrix A}
\boldsymbol{J}(\boldsymbol{x})=\left[
\begin{array}{cccc}
0
 & 1 &  \ldots  & 0  \\
\dfrac{\partial f_1}{\partial x_1}
 & \dfrac{\partial f_1}{\partial x_2}  & \ldots & \dfrac{\partial f_{1}}{\partial x_{32}}   \\
 \vdots & \vdots &  \ddots & \vdots    \\
 \dfrac{\partial f_{16}}{\partial x_{1}} & \dfrac{\partial f_{16}}{\partial x_{2}} & \ldots & \dfrac{\partial f_{16}}{\partial x_{32}} 
\end{array}
\right]_{[32\times 32]}
\end{equation}
As time $t$ ($x_{33})$ is not a physical dynamical state of the system, the last row and column of $\boldsymbol{J}(\boldsymbol{x})$ are omitted, as they only introduce a trivial zero Lyapunov exponent without affecting the physically meaningful ones. Some of the functions $f_j$ ($j=1,\ldots,16$) are non-smooth in $y_{ei}$ and thus in $\boldsymbol{x}$ leading to non-differentiability at the switching surfaces. Hence, $\partial{f_j}/\partial{\boldsymbol{x}}$ is computed analytically within each active regime, corresponding to $x_{4i} > 0$ or $x_{4i} < 0$ for $i=1,\ldots,4$. 
The grey-model parameters $\widehat{\boldsymbol{\theta}}_l$ ($l=1,\ldots,16$) and the track-irregularity parameters $\{\widehat{p}_1,\ldots,\widehat{p}_8\}$ are estimated according to \autoref{App:A} and \autoref{S.4.1}, respectively. The states ${x_1,\ldots,x_{32}}$, along with $\Ddot{x}_{wi}$, $\dot{x}_{wi}$, and $x_{wi}$ ($i=1,\ldots,4$) are derived from acceleration signal measurements as in \autoref{Signal_Derivation}. The Jacobian is then calculated at each sampled state $\boldsymbol{x}[k]$, yielding $\boldsymbol{J}[k]$, and assumed constant over each sampling interval [$kT_s,(k+1)T_s$].

\end{document}